# Unsupervised Learning for Fault Detection of HVAC Systems: An OPTICS -based Approach for Terminal Air Handling Units

Farivar Rajabi and J. J. McArthur*

Dept. Architectural Science, Toronto Metropolitan University, Canada

*corresponding author*

## 1 Abstract

The rise of AI-powered classification techniques has ushered in a new era for data-driven Fault Detection and Diagnosis (FDD) in smart building systems. While extensive research has championed supervised FDD approaches, the real-world application of unsupervised methods remains limited. Among these, cluster analysis stands out for its potential with Building Management System (BMS) data. This study introduces an unsupervised learning strategy to detect faults in terminal air handling units and their associated systems. The methodology involves pre-processing historical sensor data using Principal Component Analysis (PCA) to streamline dimensions. This is then followed by OPTICS clustering, juxtaposed against k-means for comparison. The effectiveness of the proposed strategy was gauged using several labeled datasets depicting various fault scenarios and real-world building BMS data. Results showed that OPTICS consistently surpassed k-means in accuracy across seasons. Notably, OPTICS offers a unique visualization feature for users called "reachability distance," allowing a preview of detected clusters before setting thresholds. Moreover, according to the results, while PCA is beneficial for reducing computational costs and enhancing noise reduction—thereby generally improving the clarity of cluster differentiation in reachability distance—it also has its limitations, particularly in complex fault scenarios. In such cases, PCA's dimensionality reduction may result in the loss of critical information, leading to some clusters being less discernible or entirely undetected. These overlooked clusters could be indicative of underlying faults, and their obscurity represents a significant limitation of PCA when identifying potential fault lines in intricate datasets.

## 2 Introduction

Building systems, especially those integrated with heating, ventilation, and air conditioning (HVAC), are often prone to faults. Such faults can result in higher energy usage, escalated maintenance costs, and suboptimal indoor environmental conditions. These disturbances often stem from challenges like sensor inaccuracies, equipment failures, or system operation errors. Recent studies suggest that due to these lapses in building systems and associated control discrepancies, there is a potential for energy wastage in the range of 15% to 30% [1]. Consequently, the significance of fault detection and diagnostics (FDD), also commonly termed as automated fault detection and diagnostics (AFDD), is paramount and is essential for preserving consistent system performance and minimizing energy wastage.

Over the years, a diverse range of FDD strategies has been devised. Contemporary studies have employed methods anchored in physical models, rule-based techniques, and data-driven approaches specifically for HVAC system FDD. Comprehensive reviews on this topic can be found in [1-3]. As the field of data science evolves and the adoption of building automation systems (BAS) along with other smart building technologies expands, data-driven FDD is garnering heightened attention. This is a marked departure from the traditional rule-based or expert knowledge methodologies. The allure of data-driven FDD lies in its minimal reliance on prior knowledge and its potential to offer heightened detection and diagnostic accuracy at a more cost-effective rate [4]. Although, its efficacy hinges on access to a vast reservoir of high-quality operational data, as highlighted by Yang et al. [5], emerging techniques that extract such data from building management systems (BMSs) are paving the way to overcome this limitation.

This study introduces a fault detection approach that employs cluster analysis to pinpoint potential anomalies in fan coil units (FCUs), terminal units tasked with heating and cooling rooms served by a dedicated outdoor air system. Time-series data sourced from the Building Management System (BMS) is processed using the Ordering Points to Identify the Clustering Structure (OPTICS) algorithm. When paired with Principal Component Analysis, this methodology effectively distinguishes between standard and anomalous operations. To gauge the effectiveness of the proposed method, multiple simulation-based labeled datasets alongside real-world FCU instances were examined. The findings were benchmarked against those derived from the k-means method, and the influence of PCA was meticulously explored. Moreover, all real-world data traces underwent review by a Certified Energy Manager well-versed in the building systems, ensuring accurate findings validation and comprehensive fault labeling to augment the learning process.

## 3    Literature Review

The prompt and accurate detection and diagnosis of faults within HVAC systems are paramount, not only to deter potential damage but also to ensure the continual provision of essential services. However, the landscape of FDD in this realm is fraught with intricate challenges. For instance, HVAC system operations are intricately linked to dynamic factors like occupancy and ever-changing weather conditions, often causing a blend of 'normal' and 'abnormal' operational patterns, making diagnostics challenging [6]. Further complicating the matter, these systems have a dense web of interconnections, exemplified by FCUs linked to central heating and cooling setups. Such interconnectedness can sometimes mask faults, causing one malfunction to offset another, thereby clouding the broader system's diagnostics [6]. Additionally, the vast diversity in building types, each with its unique HVAC demands, combined with the intricate couplings of various HVAC components, amplifies the FDD challenge. It's crucial, then, for contemporary FDD strategies to transcend conventional approaches, embracing a holistic view that addresses both singular and interconnected system intricacies.

### 3.1    FDD Approaches

The building research community consistently underscores the importance of FDD in HVAC systems, recognizing its pivotal role in optimizing energy efficiency and guaranteeing uninterrupted service to building occupants. HVAC systems encompass primary components like chillers, cooling towers, and boilers that deliver heating and cooling sources. Simultaneously, secondary elements like air handling units (AHUs), terminal boxes, fan coils, and ductworks are responsible for heat exchange and air circulation. A significant portion of recent research predominantly targets the detection and diagnosis of faults in AHUs and chillers [6]. Additionally, studies also concentrate on sensors within the HVAC, vapor compression systems, and Variable Refrigerant Flow systems. According to Katipamula and Brambley's classification [2] , FDD techniques for HVAC systems in buildings are categorized into three main types: qualitative, quantitative, and process history-based approaches. Quantitative model-based techniques employ physics-based models, either simplified or detailed, to identify discrepancies between the measured system conditions and expected operational norms. In contrast, qualitative model-based methods lean on expert-defined rules or foundational principles for the FDD process. On the other hand, process history-based methods adopt a data-driven approach, scrutinizing system sensor data to ascertain and diagnose the operational state of HVAC systems. In essence, both quantitative and qualitative model-based FDD methods are grounded in physics or engineering knowledge, classifying them as knowledge-based approaches. Conversely, process history-based methods, which don't require an understanding of the underlying physical process and solely depend on system sensing data, are termed as data-driven approaches.

### 3.1.1 Knowledge-based FDD

Knowledge-based FDD methods utilize physical and engineering principles, often through white-box and grey-box modeling, to identify HVAC faults. By comparing real-time HVAC operation with established baselines from these models, deviations signal potential issues. This residual analysis is applied at various scales, from whole buildings to specific HVAC subsystems. For instance, at building level, Maile et al. [7] introduced a comparison method for energy performance to spot energy usage anomalies in buildings. They used it on a campus building, comparing real data with design-goal simulations, structured by a building object hierarchy. Similarly, Lee et al. [8] explored the ASHRAE simplified energy analysis procedure for HVAC fault detection at the whole-building level. By calibrating simulated data with actual heating and cooling measurements, future consumption is predicted and compared to real data to identify deviations. At the HVAC system level, Ranade et al. [9] developed a grey-box model for diagnosing fan-coil units. They presented a simplified heat exchanger coil model based on polynomial regression for fault isolation. Dong et al. [10] utilized BIM (Building Information Modeling) to simplify the creation of physical models for the FDD procedure. Besides physics-based modeling techniques, another type of knowledge-based method is the analysis-driven approach, which leverages engineering insights into fault symptoms and system layouts for FDD. An exemplar of this is the Bayesian Network, a probabilistic graphical model that delineates variable relationships using edges and nodes. As an example, Zhao et al. utilized Diagnostic Bayesian networks to identify typical faults in AHUs, such as those in dampers, fans, filters, sensors, and coils, drawing from experimental data provided by ASHRAE Project RP-1312 [11]. In another study, to address both varied operating modes and component interconnections, Verbert et al. [12] developed a multi-model system-level methodology that utilizes distinct Bayesian networks for each mode. Another analysis-based method is the performance indicator approach. It monitors system functionality using measures developed from engineering evaluations of systems. This often involves using performance metrics, system operation breakdowns, and virtual sensors for FDD [13] . For example, Li and Braun [14] used a decoupling strategy to address multiple concurrent faults in vapor compression systems. In FDD, physics-based modeling methods typically don't rely on data with faults. These approaches primarily use fault-free data to set up and adjust their baseline models. Subsequently, these models are paired with predefined rules and thresholds for the FDD procedure. In the Bayesian network method for FDD, the construction of diagnostic networks relies on either fault-specific data or insights from engineering knowledge.

It is important to note that the fault data used in these methods might not originate from the buildings targeted for FDD. Instead, they can be sourced from past records of other structures, past experiments, expert insights, or previous experiences. A frequent limitation of physics-based modeling methods is their reliance on detailed building design and construction data or expert insights. These requirements often make the development of knowledge-based FDD techniques more labor-intensive.

### 3.1.2 Data-driven FDD

Apart from knowledge-based methods, data-driven strategies dominate FDD. These strategies bypass physics-based models and engineering expertise, concentrating on system sensor data to spot and isolate HVAC system malfunctions. Specifically, based on the usage of sensing data in FDD, data-driven methods can be further divided into various categories. The literature offers numerous data-driven FDD techniques which fall under supervised classification, semi-supervised, and unsupervised learning [15].

#### *3.1.2.1 Supervised learning methods*

Fault detection using supervised methods involves training with both fault--free operation and fault data to determine the status of incoming data. Depending on the output type, these supervised methods can be classified into either classification or regression techniques. Methods like Decision Tree (DT) [16] and Support Vector Machine (SVM) [17, 18] are classification techniques utilized to determine if the incoming data falls under the fault or fault-free category. As instances, Yan et al. [19] introduced a data-driven diagnostic method for AHUs using a decision tree approach, employing the classification and regression tree algorithm for tree formation. Li et al. [20] explored a tree-structured fault dependence Kernel, for improved fault detection and diagnosis in building cooling systems, emphasizing fault severity levels and their inter-dependence. Han et al. [21] introduced an optimized least squares support vector machine model for fault detection and diagnosis of a centrifugal chiller. Sun et al. [22] introduced a hybrid model for diagnosing refrigerant charge amount faults in variable refrigerant flow systems, combining support vector machine, wavelet de-noising, and an improved max-relevance and min-redundancy algorithm. On the other hand, regression techniques like Support Vector Regressions (SVR) [23] and Neural Networks

(NN) [24] aim to predict continuous variables that depict the system's operational state. These predictions are then contrasted with a baseline to detect potential faults. For example, Guo et al. [25] explored an optimized back propagation neural network approach for diagnosing faults in heating-mode variable refrigerant flow systems, enhancing diagnostic accuracy through data mining-based feature selection. To improve the efficiency of neural network-driven FDD and the feature extraction procedure, Eom et al. [26] utilized a strategy centered on convolutional neural networks for detecting refrigerant charge faults. Jiangyan et al. [27] proposed an energy diagnosis method for VRF systems utilizing SVR along with data mining and statistical quality control. Zhao et al. [28] presented an FDD approach for centrifugal chillers using SVR, integrated with control charts. Supervised methods are prevalent in detecting faults in HVAC systems. Yet, acquiring ample labeled fault data for model training is a challenge. This often results in class imbalance. Due to the hurdles and expenses in data labeling, these models might rely on data from older components or simulations, potentially reducing accuracy and increasing false alarms.

*3.1.2.2    Semi-supervised learning methods*

Alternatively, when there is a scarcity of labeled fault data, semi-supervised techniques may become a better choice [12]. These methods transfer unlabeled data into labeled classes by comparing the incoming data with normal operation, and updating the training set iteratively [29]. For example, Yan et al. [30] proposed a semi-supervised learning approach for FDD in AHUs, enhancing performance even with limited faulty training data. Similarly, Fan et al. [31] explored the potential of semi-supervised learning for detecting unseen AHU faults, highlighting the method's value in overcoming challenges posed by limited labeled data for efficient smart building management. In another study by Fan et al., they presented a semi-supervised neural network method for AHU fault detection. While semi-supervised techniques excel in scenarios with scarce labeled fault data, they demand more computational resources compared to supervised learning [29].

### 3.1.2.3 Unsupervised learning methods

Lastly, unsupervised methods, which don't need fault labels, can uncover hidden patterns in building datasets. Commonly used techniques include clustering algorithms and principal component analysis (PCA), often paired with pattern recognition strategies. Unsupervised methods, which rely solely on fault-free data, offer a more straightforward approach for developing and implementing fault detection systems. However, when faced with intricate fault scenarios, the accuracy and reliability of these methods may be compromised, potentially leading to misdiagnoses or missed faults. For example, Du et al. [32] applied subtractive clustering analysis to differentiate various faults in air handling units, especially emphasizing supply air temperature sensors, controller issues, and coil valve faults. This data mining approach, combined with features extracted through principal component analysis, segregated different fault categories and detected both known and new unknown faults. Narayanaswamy et al. [33] introduced Model, Cluster and Compare (MCC) algorithm that adeptly identifies anomalies by autonomously modeling, clustering, and comparing HVAC system entities. Through this approach, uncovered 78 anomalies from 237 VAV units. They also highlighted the crucial role of human involvement in refining fault detection in contemporary buildings. In another study, Novikova et al. [34] introduced a visualization-centric approach for real-time HVAC system log analysis, focusing on the clustering of stable operational patterns. By utilizing a modified density-based clustering technique, they expertly condensed vast BAS data of a three-story building into discernible patterns, represented via heat maps. Lastly, Yan et al. [35] through a combination of PCA and a density-based clustering algorithm, OPTICS, they devised an unsupervised learning approach for AHU sensor fault detection. Validated using TRNSYS simulation data, their method effectively detected both single and concurrent multiple sensor faults.

Anomaly detection algorithms rooted in cluster analysis have been effectively utilized to identify irregularities in building energy consumption patterns. By examining meter data, these methods can discern deviations from typical energy use, offering a deeper insight into potential inefficiencies or issues within the building's energy systems [36-38]. However, their application with BAS data has been infrequent and not extensively explored. Moreover, Significant research has been dedicated to larger equipment like Air-handling units (AHUs), chillers, and boilers, as highlighted by studies such as [3, 39]. In contrast, terminal units like fan coil units have seen less research focus, as noted by [40]. The intricacy of these units arises from their dependency on room occupancy and the overall effects of weather on the heating and cooling systems they connect with. Lastly, unsupervised strategies prevalent in literature often fall short of providing contextual insights before the fault detection process. For instance, density-based clustering methods like k-means and DBSCAN necessitate the pre-assignment of a specific cluster number. Such an approach is limiting because it's more prudent to understand the data landscape before categorization. Moreover, some researchers opt to fine-tune clustering algorithm hyperparameters using cluster evaluation metrics, such as the Calinski-Harabasz index [41]. Even so, there remains a challenge: a significant increase in the number of clusters may lead to an uptick in false alarms if the procedure is wholly automated. Furthermore, as HVAC systems amass vast amounts of sensor data with numerous features, visualizing created clusters via 2D plots using the PCA method can be inadequate; potentially significant data could be overlooked.

To bridge these gaps, our study employs the OPTICS clustering algorithm. Paired with the informative 'reachability distance plot', this allows users to visually comprehend distances between data points, providing a clearer prelude to threshold setting for cluster extraction. Emphasizing the merits of unsupervised strategies, our research not only applies to real building BAS data but also benchmarks against simulated labeled data. We also investigated the effect of using PCA on the omission of fault-related data by contrasting its application against scenarios without PCA.

## 3.2 OPTICS

OPTICS algorithm [42] is a density-centric clustering technique that stems from the foundational principles of the DBSCAN method. At its core, density-based clustering approaches aim to detect areas with a high concentration of data points, distinguished by regions with fewer data points. Unlike partitioning techniques such as k-means, which need a predefined number of clusters, density-based strategies like OPTICS can recognize clusters of varying shapes and are inherently resilient to noise in the data [18]. In density-based clustering methodologies, each data point within a specific cluster should have a predefined number of neighboring data points ($MinPts$) within a certain radius ($Eps$). This ensures that data points in dense areas are grouped together.

Below is a detailed breakdown of the OPTICS algorithm formulation:

For point $p$ to be directly density-reachable from point $q$, it must meet the following conditions.

$$p \in N_{Eps}(q) \ \& \ |N_{Eps}(q)| > MinPts \quad (1)$$

where $N_{Eps}(q)$ refers to the set of data points within the radius $Eps$ of point $q$. The reachability-distance between point $p$ and $o$ is expressed as follows.

Reachability- distance

$$= \begin{cases} Undefined, & if \ |N_{Eps}(o) < MinPts| \\ \max(core - distance(o), distance(o,p)), otherwise \end{cases} \quad (2)$$

While,

Core − distance

$$= \begin{cases} Undefined, \ if \ |N_{Eps}(p) < MinPts| \\ MinPts - distance(p), otherwise \end{cases} \quad (3)$$

The OPTICS algorithm begins by selecting a randomly selected point, then moves to its directly density-connected neighbors. Points are then arranged based on their reachability-distance in increasing order, and the algorithm continues from the first point that hasn't been expanded in the sequence. This approach essentially traverses the Minimum Spanning Tree [43].

## 3.3 PCA

PCA is a widely-adopted technique for dealing with datasets with many interrelated variables. It transforms the original dataset into a series of principal components (PCs). For a dataset with '$n$' dimensions, there will be '$n$' PCs. Typically, only the first few PCs are required to capture the significant variations in the data. For a dataset $X$ of dimensions $m \times n$ (with '$m$' being the sample data count and '$n$' the variable count), the PCs are derived from the eigenvectors and eigenvalues of its covariance matrix, as defined in Eq. 1 [35].

$$Cov(x) = \frac{X^T X}{(m-1)} \tag{4}$$

The primary PC corresponds to the largest eigenvalue, the next PC to the second largest, and so forth. Multiple techniques exist to decide how many PCs to retain. In this research, the scree diagram method, as outlined by Wang and Xiao [44], was employed. Once the PC count is established, the original data is transformed into this new PC space, leading to a condensed dataset that still captures the most critical data variations. An essential step is the standardization of sensor readings, given they might have different units and scales; thus, they're adjusted to have a mean of zero and a variance of one before applying PCA.

## 4 Methodology

This research assessed the effectiveness of the proposed fault detection technique using data from three different FCUs. The focus was on three units, namely FCU-XX, FCU-YY, and FCU-ZZ, located in office areas of a multi-use building within the Toronto Metropolitan University campus. The FCUs operate under a constant air volume system to maintain the designated air flow, modulating between occupied and unoccupied modes to uphold the desired zone temperature using in-slab heating or cooling as the primary temperature control stage. They feature a CO2 control for ventilation based on space occupancy levels, seasonal control valves for transitioning between radiant heating and cooling, and are monitored by the Energy Management and Control System (EMCS) for efficient performance and alarms for in-slab temperature deviations. The selection of these FCUs aimed to demonstrate the applicability and scalability of the method to a wider array of units and the corresponding systems. By examining FCUs across varied room sizes and occupancy levels, the goal was to emphasize the flexibility of the method under diverse conditions and environments. This versatility suggests a wider application to an extensive range of equipment. Furthermore, the scenarios chosen span a range of potential fault situations, allowing for a comprehensive evaluation across various fault scenarios. Table 2details the characteristics of the selected FCUs, highlighting the spaces they cater to and their related thermal zones. Notably, all these units are connected to a single central air handling system.

*Table 1 Properties of FCUs*

| FCU | Feeds | Location |
| --- | --- | --- |
| **FCU-XX** | 50 $m^2$ entrance lobby | Thermal zone 3 - level 01 |
| **FCU-YY** | 176 $m^2$ open-plan student lounge | Thermal zone 3 - level 01 |
| **FCU-ZZ** | 8 $m^2$ study room | Thermal zone 3 - level 02 |

The chosen units are typical of the majority found in the building, catering to the heating and cooling needs of offices situated on the first and second floors. Data was collected from FCU-XX, FCU-YY, and FCU-ZZ throughout the year, specifically during the height of the heating season in December 2019 and the peak cooling period in June 2020.

FCU-XX, the initial test unit, was selected to highlight the effectiveness of the proposed method in identifying FCU faults using cluster analysis and its following steps. This specific scenario was chosen because it displayed a clear fault. Initially, the data were classified into either cooling or heating operations. The fault detection then exclusively utilized data from that particular operation. In this research, data representing the HVAC mechanical cooling function was further examined to demonstrate the efficiency of the recommended fault detection technique. The following stage involved using a scree plot to ascertain the number of principal components, eventually settling on two PCs.

The proposed approach starts with pre-processing historical sensor data through PCA to reduce the number of dimensions. After this step, the data is grouped using the OPTICS clustering algorithm. The performance of proposed method is evaluated by applying it to various publicly available labeled datasets that represent different fault conditions, as well as real-world data from a multi-purpose building. These findings were then benchmarked against the outcomes from a conventional clustering method, k-means. The proposed fault detection approach for FCUs is depicted in Fig. 1. The methodology mainly involves three phases: pre-processing the data, clustering the data, and assessing the results. The unsupervised approach operates on the foundational premise that normal and faulty data exhibit distinct characteristics. As a result, these two types of data will inherently manifest in spatially separate clusters. By leveraging clustering techniques, one can effectively identify and differentiate between these distinct data groupings, highlighting potential anomalies or faults.

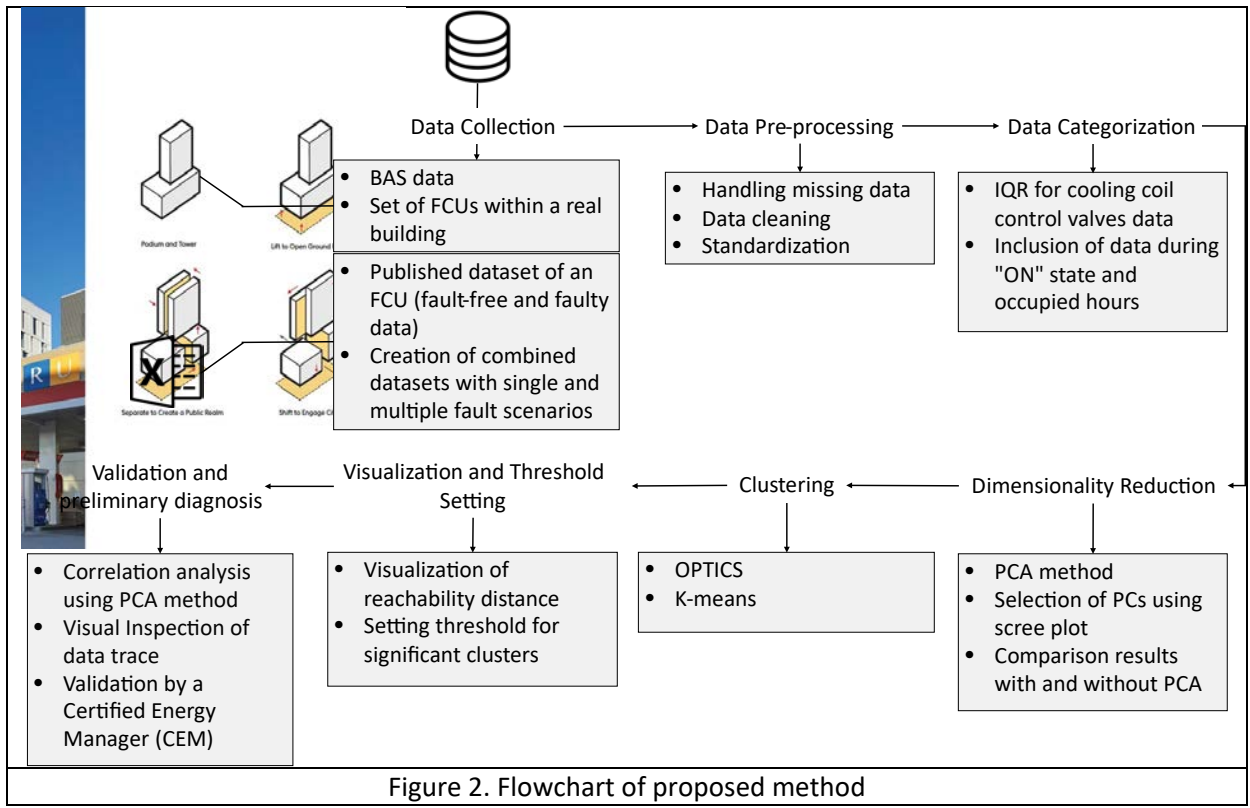

Figure 2. Flowchart of proposed method

## 4.1  *Data Collection*

This research assessed the effectiveness of the proposed fault detection technique using data from three different FCUs. The focus was on three units, namely FCU-XX, FCU-YY, and FCU-ZZ, located in office areas of a multi-use building within the Toronto Metropolitan University campus. The FCUs operate under a constant air volume system to maintain the designated air flow, modulating between occupied and unoccupied modes to uphold the desired zone temperature using in-slab heating or cooling as the primary temperature control stage. They feature a CO2 control for ventilation based on space occupancy levels, seasonal control valves for transitioning between radiant heating and cooling, and are monitored by the Energy Management and Control System (EMCS) for efficient performance and alarms for in-slab temperature deviations. The selection of these FCUs aimed to demonstrate the applicability and scalability of the method to a wider array of units and the corresponding systems. By examining FCUs across varied room sizes and occupancy levels, the goal was to emphasize the flexibility of the method under diverse conditions and environments. This versatility suggests a wider application to an extensive range of equipment. Furthermore, the scenarios chosen span a range of potential fault situations, allowing for a comprehensive evaluation across various fault scenarios. Table 2 details the characteristics of the selected FCUs, highlighting the spaces they cater to and their related thermal zones. Notably, all these units are connected to a single central air handling system.

| FCU | Feeds | Location |
| --- | --- | --- |
| FCU-XX | 50 $m^2$ entrance lobby | Thermal zone 3 - level 01 |
| FCU-YY | 176 $m^2$ open-plan student lounge | Thermal zone 3 - level 01 |
| FCU-ZZ | 8 $m^2$ study room | Thermal zone 3 - level 02 |

The chosen units are typical of the majority found in the building, catering to the heating and cooling needs of offices situated on the first and second floors. Data was collected from FCU-XX, FCU-YY, and FCU-

ZZ throughout the year, specifically during the height of the heating season in December 2019 and the peak cooling period in June 2020.

FCU-XX, the initial test unit, was selected to highlight the effectiveness of the proposed method in identifying FCU faults using cluster analysis and its following steps. This specific scenario was chosen because it displayed a clear fault. Initially, the data were classified into either cooling or heating operations. The fault detection then exclusively utilized data from that particular operation. In this research, data representing the HVAC mechanical cooling function was further examined to demonstrate the efficiency of the recommended fault detection technique. The following stage involved using a scree plot to ascertain the number of principal components, eventually settling on two PCs.

### *4.2  Data Categorization*

From the historical data sourced from the BMS, an initial categorization is undertaken to divide the data based on HVAC system operation modes, such as on/off status, heating, cooling, and natural ventilation. This is crucial as varying operation modes can lead to the formation of distinct clusters. This research specifically utilizes the dataset associated with the HVAC's mechanical cooling operation to showcase the efficacy of the suggested fault detection approach. For fault detection, operational data is utilized, specifically focusing on the cooling coil control valves' opening control signal within the interquartile range (IQR) – representing the densest observations. This stage's primary objective is to eliminate data with pronounced dynamics and data captured under extreme operational scenarios.

### *4.3  Dimensionality Reduction Using PCA*

During the subsequent stage, Principal Component Analysis (PCA) is employed. Typically, BAS data for HVAC system fault detection is packed with numerous variables, making it high-dimensional. Condensing this data can optimize computational efficiency and potentially improve clustering outcomes. However, this study evaluates the impact of both incorporating and omitting PCA prior to clustering. There are primarily two ways to reduce dimensions: feature selection and feature extraction. While the former involves selecting a subset by removing irrelevant or noisy dimensions, the latter condenses the essential information from the original data into fewer dimensions. In this context, this research employs PCA as the feature extraction method.

It's essential to highlight a distinction in this research: whereas many PCA-driven fault detection approaches use PCA for both dimensionality reduction and predicting residuals, here, PCA is solely employed for the former. This sidesteps the need for a predictive model's training and validation, typically seen in other PCA methodologies. Furthermore, the threshold concept applied here is not identical to those used in traditional PCA-based fault detection methods.

An imperative step in the analysis involves evaluating the potential benefits and limitations of dimensionality reduction, especially through PCA. To gain comprehensive insight into its impact on the clustering process, datasets are subjected to two distinct treatments: one with PCA applied and the other without any dimensionality reduction. By juxtaposing the clustering outcomes from these two strategies, we can ascertain whether PCA enhances or perhaps diminishes the clustering quality. Such a comparative analysis is pivotal in ensuring the robustness of the clustering results and discerning the most effective strategy for managing high-dimensional datasets.

### *4.4 Clustering Analysis with OPTICS Method*

Following the data pre-processing, the clustering is performed using the OPTICS algorithm [42].In the context of the OPTICS algorithm, after processing the data and producing an ordering of data points based on their density-based clustering structure, a reachability plot is generated. The y-axis of this plot represents the reachability distance, and the x-axis corresponds to the data points in the order determined by OPTICS. Fig. 3 displays a comparison of the OPTICS algorithm results on a labeled fault-free dataset (left) versus a labeled dataset containing faulty data (right). Notably, the sequence in which the data points are ordered in reachability plot, based on their traversal in the Minimum Spanning Tree, differs from their real presentation in time-series dataset. The threshold plays a crucial role in the visualization and interpretation of the reachability plot. It helps to:

**Extract Clusters**: By setting a threshold value on the reachability plot, one can visually and intuitively extract clusters. Points below the threshold line in the reachability plot are considered part of clusters, while points above the threshold often represent noise or outliers.

**Adapt to Varying Densities**: Different threshold values can highlight clusters of different densities. A lower threshold might delineate denser clusters, while a higher threshold could capture broader, less dense clusters.

**User-driven Interpretation**: The threshold can be adjusted based on user intuition. This flexibility contrasts with methods that require a predefined number of clusters or fixed parameters.

In this study, for each test case, we tailored the threshold value based on the reachability plot to extract the most significant clusters. For example, in Fig. 2 (left) a threshold value of 4.8 was chosen to extract the most significant clusters based on the reachability plot. However, there is flexibility in this approach. By reducing the threshold value, one can extract denser yet potentially less significant clusters.

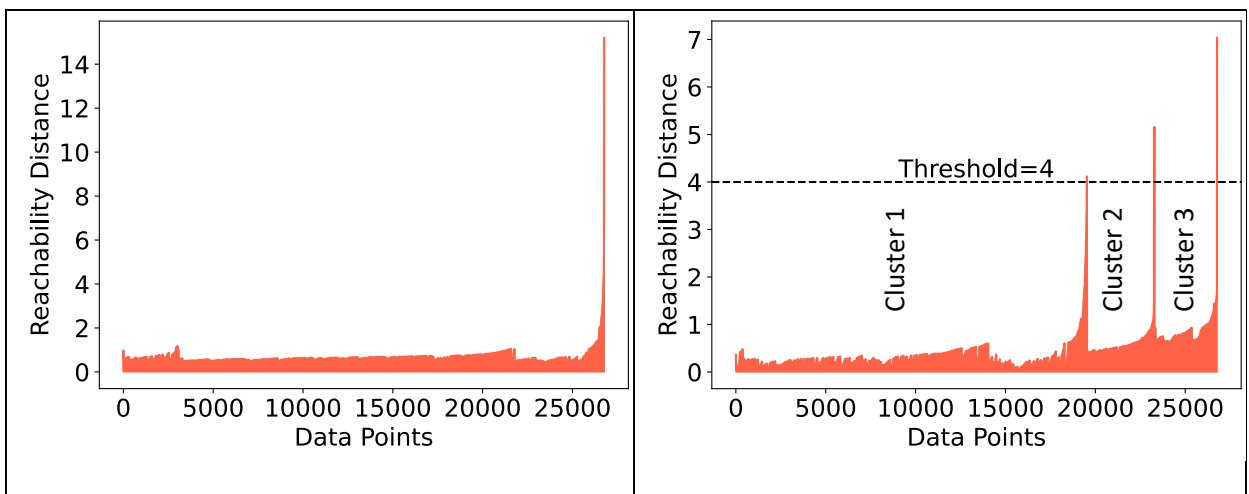

Figure 1. OPTICS analysis results for a fault-free dataset (left) and a dataset with faulty data (right)

A subsequent analysis of these clusters is imperative to determine if they represent different modes of the system's standard operation or if they indicate faults. Illustratively, in reference [35], a box-plot, a straightforward statistical method, was employed to further scrutinize the extracted clusters.

## 5 Test cases

In this research, the proposed fault detection method was evaluated using a variety of simulation-based labeled and unlabeled real-world BAS time-series datasets.

### 5.1 Published dataset: Various Labeled Fault Scenarios

The labeled dataset was sourced from a public repository [45] that documents building system operations in both faulty and fault-free conditions. This dataset encompasses the most prevalent HVAC systems and setups found in commercial buildings, spanning various climates, types of faults, and degrees of fault severity. This study utilizes an FCU designed with a vertical four-pipe hydronic setup and a fan offering three distinct speeds: high, medium, and low. The dataset covers the entire span of 2018, with observations recorded every minute with 29 features including control command and sensor measurements. More details of this case study can be found in [46]. After checking the monthly average of heating and cooling valve openings, we chose January as the heating month and June to August for cooling. For both cooling and heating months, data from the cooling coil control valves falling within the IQR was considered, while for the heating months, we focused on the data from the heating coil control valves within the IQR. Additionally, only data captured during the device's "ON" state and during occupied hours (Mon-Fri 6:00AM-6:00PM) was included. This categorization of data ensures that the presence of multiple clusters in the data points is not attributed to varying operational modes but is instead indicative of a fault. To evaluate the effectiveness of the introduced approach, a variety of fault scenarios were defined. These involve both single and multiple faults in heating and cooling months, showcasing diverse fault categories such as: heating/cooling reverse, outdoor air inlet blockage, cooling coil valve leaking, outdoor air damper stuck, cooling coil valve stuck, heating coil valve leaking, and heating coil valve stuck.

Table 1 provides details, descriptions, and timings for each test case and corresponding fault.

*Table 2. Labeled test case description*

| Type of fault(s) | Fault description | Occurring time | Purpose |
|---|---|---|---|
| **Cooling Reverse** | Cooling control reverse acting | 08-07-2018 13:49 to 08-31-2018 18:00 | Single Fault (Cooling) |
| **Outdoor Air Inlet Blockage** | Decrease damper face area | 06-04-2018 09:29 to 06-15-2018 17:08 | Single Fault (Cooling) |
| **Cooling coil valve leaking** | Assign a water flow rate when fully closed (20%, 50%, 80% of the max flow) | 06-28-2018 13:24 to 07-13-2018 15:47 | Single Fault with Different Severities (Cooling) |
| Multiple faults:<br>1- Outdoor Air Damper Stuck 80%,<br>2- Cooling Coil Valve Stuck 50% | 1- Assign a fixed simulated controlled device position<br>2- Assign a fixed simulated controlled device position | 1- 07-17-2018 06:10 to 07-20-2018 17:37,<br>2- 07-31-2018 12:19 to 08-03-2018 17:26 | Multiple Faults (Cooling) |
| Multiple faults:<br>1- Outdoor air damper stuck,<br>2- Outdoor air inlet blockage,<br>3- Heating coil valve leaking | 1- Assign a fixed simulated controlled device position,<br>2- Decrease damper face area,<br>3- Assign a water flow rate when fully closed | 1- 01-03-2018 17:22 to 01-04-2018 06:01,<br>2- 01-04-2018 06:19 to 01-04-2018 06:27,<br>3- 01-15-2018 09:49 to 01-17-2018 17:02 & 01-26-2018 08:54 to 01-26-2018 17:54 | Multiple Faults (Heating) |

To accommodate the detection of single, multiple, and non-simultaneous faults, we set the threshold values in OPTICS to ensure the extraction of the most significant clusters. Although there is not a universally accepted definition of 'significant clusters', they can typically be discerned by examining reachability plots. The most notable clusters are those valleys below the threshold, delineated by points where the reachability distance exhibits a considerable rise. This intersection between the threshold and the reachability distance is indicative of the presence of clusters.

In the application of the k-means algorithm, two approaches can be considered. The first involves determining the number of clusters by using a cluster evaluation criterion, such as the Calinski-Harabasz index [41], with the largest cluster being identified as normal operation and smaller ones as faults. This is under the assumption that faults are rare in system operation data. The second approach sets the number of clusters at two, one indicating normal operation and another indicating faulty operation. Based on our analysis, the latter approach seems to achieve better results in this context. However, the former method, which requires further manual analysis of the generated clusters as exemplified in [47], falls outside the scope of this study. Therefore, in this case the number of clusters in k-means was set as two.

## 5.2  Real-world time series data of a series of FCUs

In order to demonstrate the usefulness of proposed method with real-world BAS data, several time-series datasets have been compiled. One of these datasets had a known fault for testing purposes. data was gathered from various equipment measurements and control points within the BAS, including control commands for heating and cooling as well as various sensor readings such as room temperature, discharge air, and supply water temperature.

Table 2 provides the names of the sensors and their respective descriptions. This case study aims to demonstrate the effectiveness of our suggested fault detection technique by examining an HVAC dataset focused on mechanical cooling operations using FCUs interfaced with in-slab heating and cooling systems. The data from cooling coil control valve that fall within the IQR, representing the most consistent observations was used. Given that FCUs usually have two main operational states - cooling and heating, it is vital to differentiate between them to ensure accurate results. Moreover, the data only from periods when the device was actively turned on ("ON") was considered.

*Table 3. Names and descriptions of sensors*

| Designation | Description |
| --- | --- |
| T | Zone temperature |
| Q | Indoor air quality |
| INSLAB-T | In-slab temperature |
| DA-T | Discharge air temperature |
| CLG-O | Cooling coil valve |
| HTG-O | Heating coil valve |
| ST | Supply water temperature |
| RT | Return water temperature |
| VR | Volume rate |
| EA-T | Exhaust air temperature |
| EA-F | Exhaust air flow |
| DA-F | Discharge air flow |

In terms of hyperparameters of OPTICS, for this case, assuming single and non-simultaneous faults, it was expected that the data would form two separate clusters, representing normal and malfunctioning system operations. So, thresholds were set in the OPTICS algorithm to yield two distinct clusters. Likewise, in k-means, the number of clusters was set as two.

There is no general method for setting $MinPts$. Assigning a value that is too low could mistakenly label low-density areas with few noise points as clusters. Conversely, setting $MinPts$ too high may risk missing out on valid clusters with fewer data points. Based on [42], a range of 10-20 is recommended for $MinPts$. For this particular study, a value of 15 was selected for $MinPts$. It is worth noting that the appropriateness of different values might differ based on the dataset's attributes and the particular application's needs. Subsequently, with a specified $MinPts$, the $Eps$ value can be ascertained using the k-distance graph, following the approach outlined by Ester et al. [17]. The k-distance refers to the distance between a specific point and its $k$th closest neighboring point, with $k$ being set to $MinPts$. As illustrated in Fig. 3, the k-distance graph displays the ordered k-distances for all relevant data points. The optimal $Eps$ value is identified at the point where there is a notable increase in the k-distance. However, As indicated in [35], the performance of OPTICS is not heavily influenced by the specific choices of $Eps$ and $MinPts$.

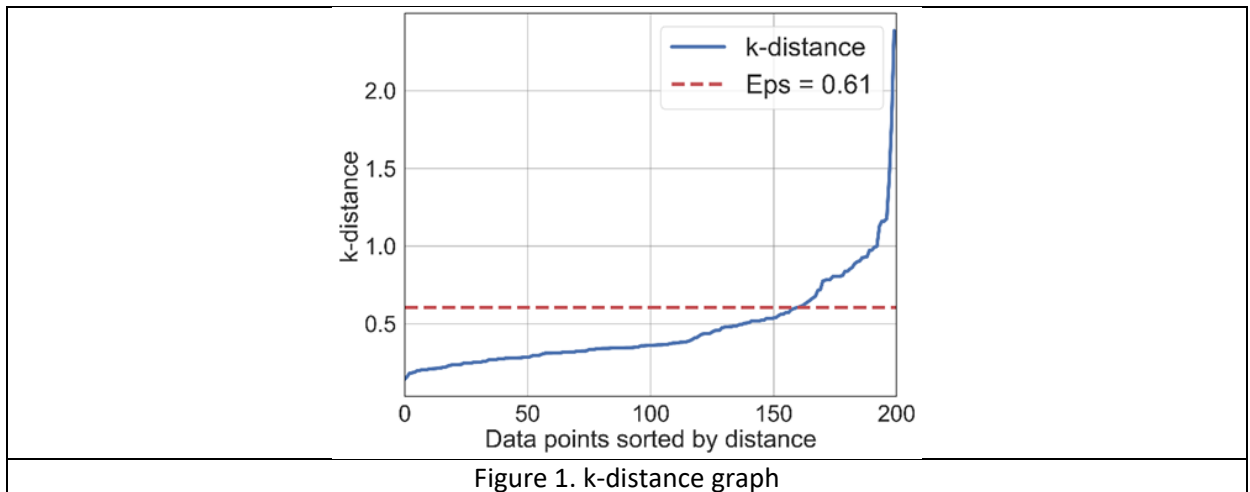

Figure 1. k-distance graph

Through examining the outcomes of the clustering and monitoring data traces from the system's operations, it is possible to pinpoint time periods of malfunction. These are discernible because they are categorized under a label that is separate from that of normal operations. Nevertheless, to delve into the likelihood of multiple faults potentially leading to the formation of multiple clusters, one could opt for a lower threshold in the OPTICS algorithm or a higher '$k$' value in k-means clustering.

For fault detection, clusters were labeled as 'normal' and 'faulty' operation, while non-clusters were marked as 'noise'. PCA plots were generated to compare the findings of OPTICS and k-means in 2D plots. These clustering outcomes were also plotted against operational data traces for a detailed examination, offering a more profound insight into the data. A Certified Energy Manager (CEM) reviewed these to identify the faults. This review had a dual purpose: it not only provided expert verification of the results beyond just the k-means comparison but also facilitated the identification of the most probable fault. This identified fault was then labeled on the faulty data, enabling the learning of diagnostic techniques in future research.

## 6 Results

In order to evaluate the reliability of the proposed fault detection method, which utilizes OPTICS clustering and PCA, an initial test was conducted on five different fault scenarios generated from a published dataset. This assessment involved a comparison between results obtained with and without the application of PCA, to determine its effect on data clarity and cluster separation. Furthermore, this phase provided a benchmark by contrasting the method's performance with that of the k-means clustering technique. Progressing from simulated environments to actual operational contexts, a subsequent exploration applies the method to field data from a building's energy management system. Here, the aim is to verify the method's effectiveness and robustness when confronted with the complexities of live, multi-variate data streams.

### 6.1 Published dataset: Various Labeled Fault Scenarios

To demonstrate the versatility of the proposed method across various fault types and situations, it was tested under multiple conditions. The outcomes were then compared with k-means, and the impact of PCA was explored for each scenario. The results are displayed in Table 4.

According to the table, in Case 1, cooling reverse fault, the performance difference between the OPTICS algorithm and k-means in detecting faults was found to be minimal, regardless of whether PCA was implemented. Both methods effectively detected faults within the timeframe from July 8, 2018 to August 31, 2018.

Variations in True Positives (TP) and False Negatives (FN) were observed, which can be attributed to the sensitivity of OPTICS to its user-defined threshold value. This sensitivity particularly affects noise detection. For the purposes of this analysis, noise identified by OPTICS was classified as fault operation.

In Case 2, outdoor air inlet blockage, a distinct variation in the performance of the OPTICS algorithm was observed when combined with PCA. Specifically, the application of OPTICS with PCA failed to detect the entire fault period from June 4, 2018, to June 15, 2018. In contrast, when PCA was not applied, OPTICS exhibited a markedly improved capability, identifying almost the entire duration of the fault.

Comparatively, the sensitivity of k-means clustering to the presence or absence of PCA was less pronounced. However, its overall performance was somewhat enhanced when PCA was not utilized. Notably, when comparing OPTICS without PCA to k-means, OPTICS demonstrated superior fault detection accuracy. This is evidenced by the observation that k-means classified approximately half of the faulty period as normal operation and misidentified several intervals of normal operation as faults, leading to a significant increase in both false positives and false negatives.

According to the table, it can be seen that OPTICS without PCA surpassed k-means without PCA by over 50% in accuracy. This substantial difference underscores the effectiveness of the OPTICS algorithm in detecting HVAC system faults under the given conditions, particularly when PCA is not applied as a preprocessing step.

In case 3, a key finding was that all approaches successfully detected the cooling coil valve leaking fault. However, standalone OPTICS distinguished itself with a superior performance. In terms of recall, standalone OPTICS was approximately 11% more effective than k-means. Regarding the influence of PCA, k-means demonstrated no significant sensitivity to the application of PCA. On the other hand, the application of PCA to OPTICS led to the algorithm labeling more data points as noise. Notably, a portion of these data points corresponded to normal operation, leading to a slight increase in the false positive rate for OPTICS with PCA..

*Table 4. Benchmark test cases results*

| Case no. | Type of Fault(s) | Algorithm | PCs | TP | FP | FN | TN | Precision | Recall | F1-score | Accuracy |
|---|---|---|---|---|---|---|---|---|---|---|---|
| 1 | Cooling reverse | OPTICS | 3 | 7219 | 0 | 9 | 19547 | 1 | 0.999 | 0.999 | 0.999 |
|   |   |   | - | 7221 | 0 | 7 | 19547 | 1 | 0.999 | 0.999 | 0.999 |
| | | k-means | 3 | 7221 | 0 | 7 | 19547 | 1 | 0.999 | 0.999 | 0.999 |
| | | | - | **7222** | **0** | **6** | **19547** | **1** | **0.999** | **0.999** | **0.999** |
| 2 | Outdoor air Inlet blockage | OPTICS | 3 | 0 | 46 | 4279 | 22450 | 0 | 0 | NaN | 0.838 |
| | | | - | **4277** | **0** | **2** | **22496** | **1** | **0.999** | **0.999** | **0.999** |
| | | k-means | 3 | 2161 | 10862 | 2118 | 11634 | 0.165 | 0.505 | 0.249 | 0.515 |
| | | | - | 2375 | 10801 | 1904 | 11695 | 0.180 | 0.555 | 0.272 | 0.525 |
| 3 | Cooling coil valve leaking | OPTICS | 3 | 3843 | 228 | 134 | 22520 | 0.944 | 0.966 | 0.955 | 0.986 |
| | | | - | **3877** | **96** | **150** | **22652** | **0.976** | **0.963** | **0.969** | **0.991** |
| | | k-means | 3 | 3443 | 0 | 584 | 22748 | 1 | 0.854 | 0.921 | 0.978 |
| | | | - | 3447 | 0 | 580 | 22748 | 1 | 0.855 | 0.922 | 0.978 |
| 4 | Cooling Coil Valve Stuck & Outdoor air damper stuck | OPTICS | **3** | **3831** | **30** | **0** | **22914** | **0.992** | **1** | **0.996** | **0.999** |
| | | | - | 3831 | 45 | 0 | 22899 | 0.988 | 1 | 0.994 | 0.998 |
| | | k-means | 3 | 1995 | 0 | 1836 | 22944 | 1 | 0.520 | 0.684 | 0.931 |
| | | | - | 1995 | 0 | 1836 | 22944 | 1 | 0.520 | 0.684 | 0.931 |
| 5 | Outdoor air damper stuck, Outdoor air inlet blockage, Heating coil valve leaking | OPTICS | 3 | 1417 | 2 | 1029 | 9776 | 0.999 | 0.579 | 0.733 | 0.916 |
| | | | - | **2446** | **46** | **0** | **9732** | **0.982** | **1** | **0.991** | **0.996** |
| | | k-means | 3 | 1200 | 0 | 1246 | 9778 | 1 | 0.491 | 0.658 | 0.898 |
| | | | - | 1364 | 0 | 1082 | 9778 | 1 | 0.558 | 0.716 | 0.911 |

In Case 4, the dataset comprised multiple, non-simultaneous faults, presenting a complex scenario for fault detection. The application of PCA to OPTICS algorithm yielded a marginally improved performance. Notably, both variants of OPTICS (with and without PCA) demonstrated superior performance compared to the k-means clustering algorithm. A critical observation was the ability of OPTICS, in both its PCA-applied and non-PCA forms, to efficiently identify two distinct periods of faults: an outdoor air damper fault occurring from July 17 to 20, and a cooling coil valve fault from July 31 to August 3. However, a slight distinction was observed in the rate of FP. The OPTICS variant without PCA registered a higher count of data points labeled as noise, leading to a slight increase in false positives.

Conversely, the k-means algorithm, while effectively identifying the cooling coil valve fault (July 31 to August 3), failed to detect the outdoor air damper fault (July 17 to 20). This resulted in a notable increase in FN for the k-means algorithm in both variant.

Case 5 involved a dataset characterized by three distinct fault scenarios: an outdoor air damper stuck (January 3-4), an outdoor air inlet blockage (January 15-17), and a heating coil valve leak (briefly on January 26). This case provided a comprehensive assessment of the fault detection capabilities of both OPTICS and k-means, with and without PCA. A key observation was the effectiveness of all algorithmic configurations in detecting the outdoor air inlet blockage and heating coil valve leaking faults, particularly in the latter two periods of faulty operation. Notably, standalone OPTICS exhibited a marginally superior performance in identifying the outdoor air inlet blockage. However, the first fault scenario, the outdoor air damper stuck, presented a notable challenge. K-means, irrespective of PCA application, failed to detect this fault. In contrast, both OPTICS variants identified this fault period, albeit primarily as noise. Standalone OPTICS labeled more data points as noise compared to OPTICS with PCA, indicating a higher sensitivity to this particular fault type. In terms of overall performance metrics, standalone OPTICS demonstrated notably better results than OPTICS with PCA. The recall rate for standalone OPTICS was 42% higher than that for OPTICS-PCA, and its F1 score was approximately 25% better. However, it is important to note that standalone OPTICS, while showing better fault detection capability, also marked more points of normal operation as noise compared to its PCA-applied counterpart. This resulted in an increased false positive rate for standalone OPTICS.

## 6.2  *Real-world time series data of a series of FCUs*

To ascertain the proposed method's performance with real-world Building Automation System data, the study scrutinized three distinct fan coil units—FCU-XX, FCU-YY, and FCU-ZZ—located within a mixed-use building.

### 6.2.1 FCU-XX

Confirmed by the facility's engineer and CEM, specific periods of malfunctions in the FCU-XX system have been identified. These malfunctions occurred during distinct intervals: from December 23, 2019, at 17:20 to December 28, 2019, at 05:20, followed by another from December 28, 2019, at 07:40 to December 31, 2019, at 12:50. The system also encountered operational faults from June 3, 2020, at 12:00 to 18:00, and a prolonged fault from June 12, 2020, at 03:20 to June 18, 2020, at 09:50. These documented fault periods provide critical benchmarks for evaluating the accuracy and reliability of the fault detection algorithms applied in this study.

To adjust the hyperparameters of the OPTICS algorithm, k-distance diagram was drawn, leading to the identification of $Eps$ as the curve's inflection point, seen in Fig. 3. This $Eps$ value was consistently 0.61 for every FCU analyzed. When applying $Eps$ to the OPTICS algorithm, the reachability distance graph was plotted. A threshold was chosen to extract the most significant clusters, signifying the system's normal and faulty operations. For FCU-XX, the threshold values were determined to be 4 in December (Fig. 4, left) and 3 in June (Fig. 4, right).

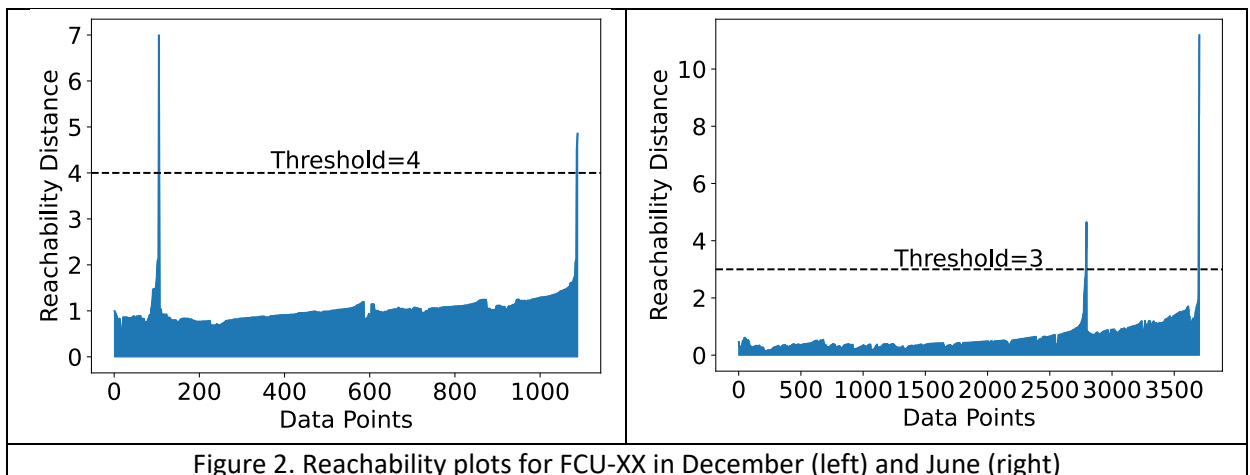

Figure 2. Reachability plots for FCU-XX in December (left) and June (right)

Based on the analysis, the data were categorized as either 'faulty' or 'normal' operation. When visualized on a time-series chart with color coding, the periods of malfunction were easily identifiable. Upon verification with the facility engineer, the identified fault was confirmed.

In December, both the OPTICS and k-means algorithms demonstrated efficiency in detecting the specified faulty periods within the FCU-XX system. The application of PCA did not significantly influence the performance of either algorithm. Both OPTICS and its PCA variant incorrectly identified a brief interval on December 9, from 11:20 to 11:40, as noise. Furthermore, all algorithms, in both their standard and PCA-applied forms, successfully detected the fault period spanning from December 23 at 17:20 to December 31 at 00:50. However, a specific timeframe within the known fault period, from December 28 at 17:20 to 19:10, was erroneously flagged as a fault by all algorithmic approaches. This misclassification contributed to a slight increase in the FP rate across all algorithms.

In the June fault detection analysis, all algorithms successfully identified the fault that occurred on June 3, from 12:00 to 18:00. However, the performance varied significantly for the fault from June 12 at 3:20 to June 18 at 9:50. Standalone OPTICS demonstrated a notably superior performance in this instance, efficiently detecting the entire duration of this fault. It is noteworthy, however, that this algorithm incorrectly classified the period from June 12, 2:10 to 3:10, as a fault, indicating a minor increase in its FP rate. Additionally, both OPTICS and its PCA variant incorrectly identified the time span from June 18, 10:00 to 16:00, as noise. Furthermore, OPTICS with PCA failed to detect the initial part of the fault, specifically from June 12 at 3:20 to June 13 at 4:30.

Regarding the k-means algorithm, the impact of PCA was less pronounced. Nonetheless, both variants of k-means missed the initial phase of the fault. The standard k-means algorithm failed to detect the fault from June 12 at 3:20 to June 13 at 1:20, while k-means with PCA extended this miss until June 13 at 2:40. The comparative outcomes of the OPTICS and k-means algorithms, with and without the implementation of PCA for the December and June datasets, are comprehensively summarized in Table 5.

Table 5. Real-world test case results

| | Month | Algorithm | PCs | TP | FP | FN | TN | Precision | Recall | F1-score | Accuracy |
|---|---|---|---|---|---|---|---|---|---|---|---|
| FCU-XX | December | OPTICS | 3 | 987 | 17 | 0 | 85 | 0.983 | 1 | 0.991 | 0.984 |
| | | | - | 987 | 16 | 0 | 86 | 0.984 | 1 | 0.992 | 0.985 |
| | | k-means | 3 | 987 | 13 | 0 | 89 | 0.987 | 1 | 0.993 | 0.988 |
| | | | - | 987 | 13 | 0 | 89 | 0.987 | 1 | 0.993 | 0.988 |
| | June | OPTICS | 3 | 792 | 42 | 149 | 2720 | 0.945 | 0.841 | 0.892 | 0.948 |
| | | | **-** | **941** | **48** | **0** | **2714** | **0.951** | **1** | **0.975** | **0.987** |
| | | k-means | 3 | 800 | 21 | 141 | 2741 | 0.974 | 0.850 | 0.908 | 0.956 |
| | | | - | 810 | 17 | 131 | 2745 | 0.979 | 0.861 | 0.916 | 0.960 |

The comparative analysis of algorithm performances, as delineated in Table 5, reveals distinct patterns in the December and June datasets. In the December dataset, the results indicate no significant difference in the performance of the algorithms, whether standalone or with the application of PCA. Contrastingly, the June dataset presents a different scenario. Here, standalone OPTICS notably outperformed the other approaches. This superiority is particularly evident in terms of Recall, where OPTICS without PCA surpassed its PCA-applied variant by approximately 15%. Moreover, standalone OPTICS also exhibited enhanced performance in Recall by about 14% compared to k-means.

### 6.2.2 FCU-YY

To assess the efficacy of the proposed fault detection method in real-world scenarios, particularly in systems without pre-identified faults, the procedure was extended to two additional FCUs, designated as FCU-YY and FCU-ZZ. These units, located within a functioning building, provided unlabeled datasets that represent typical operational environments. This application aimed to evaluate the robustness and adaptability of the method in detecting potential faults in real-time operational conditions, without the guidance of previously labeled fault data.

In the analysis of the FCU-YY dataset, the reachability plots generated by the OPTICS algorithm played a crucial role in identifying distinct operational states. By examining these plots, thresholds were determined to effectively segregate the data into clear clusters, indicative of different operational conditions or potential faults. As illustrated in Figure 5, for the December dataset (shown in the left panel of Fig. 5), a threshold of 0.7 was selected. This threshold facilitated the distinction of two prominent clusters within the data. Similarly, for the June dataset (depicted in the right panel of Fig. 5), a higher threshold value of 2 was chosen.

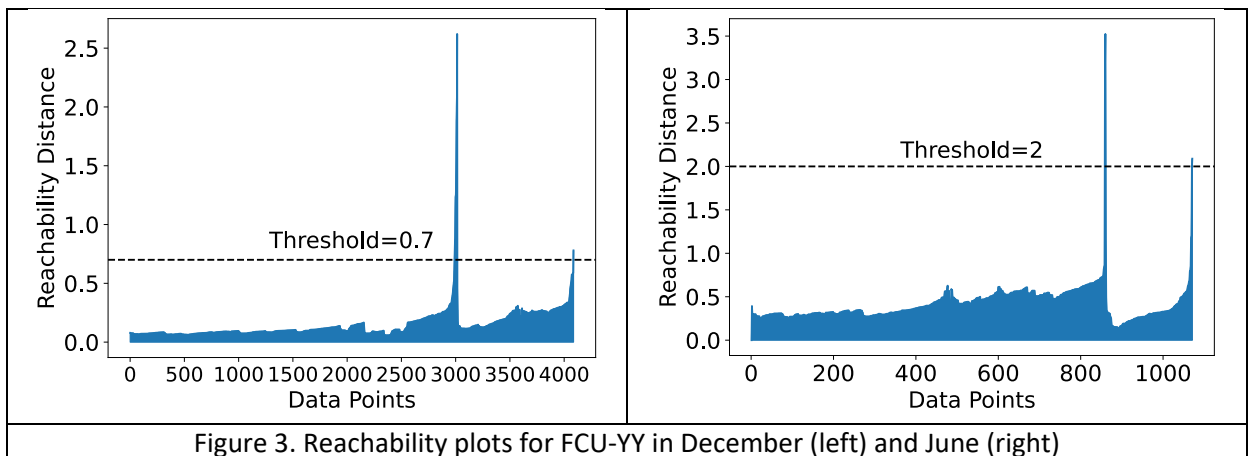

Figure 3. Reachability plots for FCU-YY in December (left) and June (right)

In the analysis of the December dataset for FCU-YY, all employed algorithms were able to detect a fault occurring from December 23 at 13:30 to December 31 at 00:50. Notably, standalone OPTICS identified an additional fault period that was not detected by OPTICS-PCA. This period spanned from December 1 at

9:50 to December 3 at 11:30. However, the application of PCA on k-means showed no significant impact on its performance. K-means detected a portion of this fault, specifically from December 2 at 23:40 to December 3 at 11:30.

Upon consultation with the Certified Energy Manager (CEM), the presence of these faults within the December dataset was confirmed. However, it was noted that the system was actually operating normally during part of the period identified by standalone OPTICS as a fault, specifically from December 1 at 9:50 to 11:30. This misclassification by OPTICS indicates a false positive detection, erroneously interpreting normal operation as a fault (noise).

In the analysis of the June dataset for FCU-YY, standalone OPTICS demonstrated a notable detection capability by identifying a fault period from June 12 at 11:00 to June 18 at 13:30. In contrast, OPTICS with PCA detected only a portion of this fault, specifically from June 13 at 14:10 to June 18 at 13:30. The application of PCA on k-means did not show a significant impact, as k-means in both its standard and PCA-applied variants identified the fault from June 13 at 7:40 to June 18 at 10:00. Following consultation with the CEM, it was confirmed that a fault indeed occurred from June 12 at 11:00 to June 18 at 10:00. This confirmation underscores that while standalone OPTICS was the most effective in detecting the entire duration of the fault, it also erroneously classified a small time segment from June 18 at 10:10 to 13:30 as noise.

### 6.2.3 FCU-ZZ

In the analysis of FCU-ZZ by examining reachability plots, thresholds were determined. As illustrated in Figure 6, for the December dataset (shown in the left panel of Fig. 6), a threshold value of 1 was chosen. Similarly, for the June dataset (depicted in the right panel of Fig. 6), a threshold of 4 was selected.

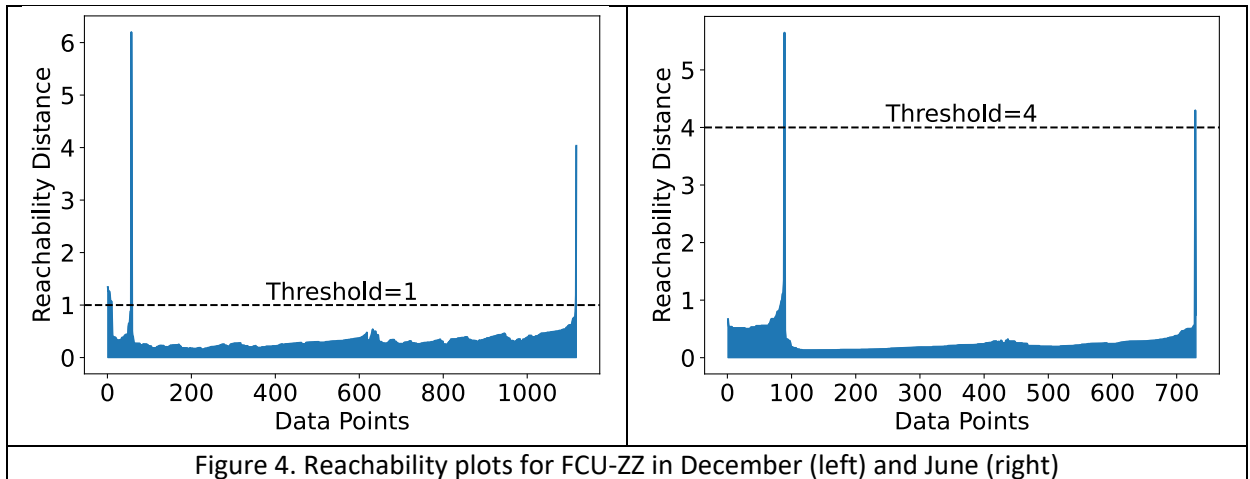

Figure 4. Reachability plots for FCU-ZZ in December (left) and June (right)

In the December analysis of the FCU-ZZ dataset, all the algorithms consistently detected a fault occurring from December 23 at 14:40 to December 31 at 00:50. Upon consultation with the CEM, the occurrence of this fault during the specified period was confirmed. However, it was also brought to attention by the CEM that an additional fault occurred from December 2 at 14:00 to 21:00. This fault was not identified by any of the algorithms used in the study.

In the analysis of the June dataset for FCU-ZZ, all algorithms successfully identified a fault occurring from June 1 at 9:20 to June 3 at 18:20. The CEM confirmed the existence of this fault during the specified timeframe, validating the algorithms' detection accuracy for this period.

However, standalone OPTICS identified an additional fault period from June 3 at 18:30 to 19:10, which was subsequently confirmed by the CEM to be a false detection, incorrectly categorizing this time as a fault (noise). Further, standalone OPTICS detected another fault from June 12 at 9:50 to June 18 at 10:00. OPTICS with PCA also detected a portion of this fault period, from June 13 at 16:40 to June 18 at 10:00. Similarly, k-means, both with and without PCA, identified a nearly similar fault period from June 13 at 10:50 to June 18 at 10:00. Consultation with the CEM revealed that standalone OPTICS accurately detected the entire duration of this fault, while the other algorithms missed the initial part of the fault period.

### 6.2.4 Preliminary fault diagnosis

To initiate a preliminary diagnosis of faults within the system, PCA was conducted to quantify the degrees of interdependency among various features. Table 6 presents the weights for each feature on the first two principal components (PC1 and PC2) obtained PCA analysis. This analysis reveals the degree of influence each feature has on the directions of maximum variance in the dataset. The loadings indicate how much each original feature contributes to the principal components, providing insights into the underlying structure of the dataset.

*Table 6 Principal Components for FCU-XX*

| Feature | PC1 weight | PC2 weight |
|---|---|---|
| T | 0.61 | 0.46 |
| Q | -0.06 | -0.11 |
| INSLAB-T | 0.52 | 0.45 |
| DA-T | 0.76 | 0.19 |
| CLG-O | -0.75 | 0.13 |
| HTG-O | 0.61 | 0.27 |
| ST | 0.45 | 0.68 |
| RT | 0.42 | 0.57 |
| VR | -0.21 | 0.41 |
| EA-T | -0.36 | 0.53 |
| EA-F | -0.26 | 0.33 |
| DA-F | -0.10 | 0.18 |

As observed in Table 6, when variables change in the same direction, such as T, INSLAB-T, and HTG-O, there is a direct correlation. Conversely, when there's a change in opposite directions, as seen between T and CLG-O, an inverse correlation exists. However, when both principal components exhibit opposing shifts, like T and Q, no correlation is observed. This analytical process revealed a pronounced correlation between the cooling valve opening (CLG-O) and the discharge air temperature (DA-T). Under normal operational conditions, these two parameters are expected to exhibit a directly proportional relationship, altering in tandem along a predictable diagonal trajectory. Any deviation from this trajectory would suggest an anomaly, potentially indicative of a fault, prompting further investigation into system performance and behavior.

To delve deeper into the cause of the faults, the periods identified as faulty by the proposed method were highlighted with boxes on the data traces for the three FCUs in Fig 7. These visual markers on the time-series data enable a focused examination of the system's behavior during the flagged intervals. By scrutinizing these segments, insights can be gleaned into the discrepancies between expected and actual performance, thus facilitating the diagnosis of potential fault roots. This visualization also provides a clear representation of the identified malfunction periods, offering insights into the temporal distribution and duration of faults as detected by the algorithm.

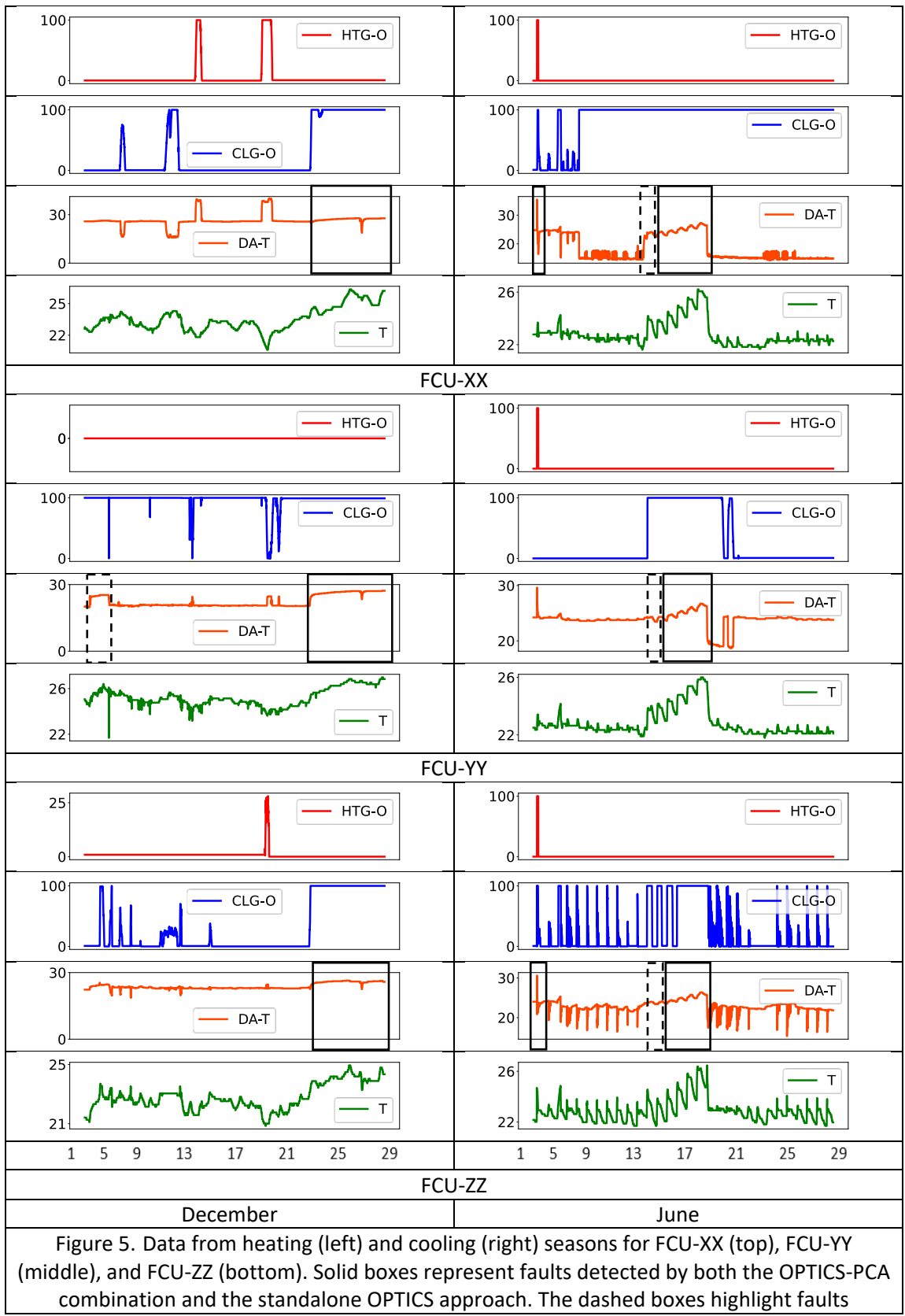

Figure 5. Data from heating (left) and cooling (right) seasons for FCU-XX (top), FCU-YY (middle), and FCU-ZZ (bottom). Solid boxes represent faults detected by both the OPTICS-PCA combination and the standalone OPTICS approach. The dashed boxes highlight faults

identified exclusively by the standalone OPTICS method, which were not detected when OPTICS was used in conjunction with PCA.

According to the visual cues presented in Fig 7, it was observed that during specific intervals in December and June, the Discharge Air Temperature (DA-T) failed to respond appropriately to the Cooling Output (CLG-O). Notably, in the instances marked on the figure, even when CLG-O was fully operational at 100%, there was no corresponding decrease in DA-T, which anomalously remained elevated. This pattern was consistently observed across all three FCUs investigated. A critical analysis of this trend by the CEM led to the hypothesis of a systemic issue within the central plant system as in these time periods, there is insufficient chilled water flow to fulfill the cooling needs. This diagnosis implies that the malfunction was not due to local unit discrepancies but likely stemmed from a broader, centralized source affecting multiple FCUs concurrently.

# 7 Discussion

…In this section, a comparative analysis is presented between the OPTICS and k-means clustering algorithms, focusing on their performance in various fault scenarios. The effect of PCA on these algorithms is also evaluated, aiming to ascertain its impact on detection capabilities. This analysis seeks to elucidate the relative strengths and limitations of each algorithm under different conditions.

## 7.1 Comparison between OPTICS and k-means

A distinct advantage of OPTICS is its ability to offer users a compelling visualization tool in the form of a 'reachability plot'. This visualization not only provides insights into the data structure but also empowers users to set thresholds intuitively. By examining the reachability plot, users can make informed decisions on where to draw thresholds, separating the most significant clusters based on the data's inherent characteristics rather than arbitrary decisions. This is especially advantageous when compared to algorithms like k-means that often rely on setting a numerical threshold for cluster separateness. Such thresholds can sometimes be arbitrary and not truly representative of the underlying data structure. For example, the k-means algorithm often uses the silhouette index to determine the optimal number of clusters, requiring users to set a threshold that might not always capture the nuances of the data. OPTICS, with its visually informed threshold-setting capability, offers a more grounded and reasonable approach, ensuring that the clustering genuinely reflects the intrinsic patterns and variances within the data.

Further setting OPTICS apart from traditional algorithms like k-means is its ability to detect clusters of varying shapes, going beyond the conventional spherical or circular patterns typical of k-means. Its density-based nature also grants it a robustness against noise, ensuring that it remains less influenced by outliers which might skew cluster formation in other algorithms. Moreover, a salient advantage of OPTICS is its independence from pre-defined cluster counts. Instead of relying on preset values, it determines the number of clusters based on the data's intrinsic density, ensuring a more organic and accurate data representation.

The combined strengths of OPTICS—its ability to form clusters of diverse shapes, its resilience against noise, and the lack of a need for pre-specifying cluster counts—offer a nuanced data representation that is invaluable for professionals in fault detection and diagnosis. Such capabilities ensure that anomalies or faults identified are genuine reflections of system issues, rather than mere artifacts of the data or clustering process.

## 7.2 Effect of PCA

Utilizing Principal Component Analysis (PCA) offers advantages in processing multidimensional data. By reducing irrelevant dimensions, PCA enhances the identification of genuine clusters by both minimizing noise and amplifying the contrast between clusters. When PCA is not applied, increasing data dimensions can result in data sparsity. As the dimensions grow, similarity measures such as Euclidean distance may become less meaningful due to the increased dispersion of data points in the space. However, while PCA aids in simplifying and highlighting major data patterns, it is not without risks; specifically, PCA might inadvertently omit certain clusters. These missing clusters, in some scenarios, could represent pivotal features or potential faults, implying that PCA might not capture all nuanced changes in the data. Recognizing that PCA offers distinct advantages but also comes with potential pitfalls, both scenarios are examined: one with the application of PCA and the other without. This approach provides a comprehensive assessment of PCA's influence on the clustering results, effectively capturing its strengths and potential shortcomings.

In situations where faults induce significant deviations across a wide array of features, as exemplified by Case 1 (cooling reverse), Case 4 (cooling coil valve stuck and outdoor air damper stuck) in the published dataset, and in real-world systems like FCU-XX and FCU-ZZ during December, the effectiveness of Principal Component Analysis (PCA) in streamlining the data dimensionality without impeding fault detection capabilities becomes evident. PCA's role in simplifying complex datasets, while retaining essential information for identifying faults, has proven particularly beneficial. This is notable in scenarios with intricate and overlapping fault signatures, where PCA aids in distilling key features, thereby enhancing the efficiency of the fault detection process. The principal components, especially the first few, tend to encapsulate the majority of the variance in the data, which often includes the significant deviations caused by faults. Thus, in such clear-cut cases, PCA does not impede the fault detection capabilities of the subsequent OPTICS clustering, maintaining a level of performance that is often comparable to scenarios where PCA is not employed.

In contrast, scenarios characterized by subtle, complex, or sparse deviations among features, such as Case 2 (outdoor air inlet blockage), Case 3 (cooling coil valve leaking), and Case 5 (outdoor air damper stuck, outdoor air inlet blockage, and heating coil valve leaking) in the published dataset, as well as in real-world systems like FCU-XX, FCU-YY, and FCU-ZZ during June and FCU-YY in December, have illustrated the advantages of foregoing PCA. In these instances, the direct analysis of the full feature set, without the dimensionality reduction offered by PCA, has proven more effective. The omission of PCA allows for a finer resolution in detecting nuanced or less pronounced feature variations, which is crucial in accurately identifying these more intricate fault patterns. This approach underscores the importance of a tailored analysis strategy in fault detection, where the decision to use or omit PCA is contingent upon the specific characteristics of the fault scenario. The intrinsic nature of PCA to prioritize variance can inadvertently marginalize features that, while not contributing substantially to overall variance, may be pivotal in detecting certain faults. This is particularly evident in scenarios involving multiple faults or complex fault patterns, where nuanced changes across various features may be diluted or lost post-PCA, thereby diminishing the fault detection capability of the model.

# 8 Conclusions

This paper has presented an innovative approach to fault detection and preliminary diagnosis in terminal air handling units, utilizing the OPTICS clustering algorithm. The methodology relies on the spatial separation of monitored data for fault identification. The effectiveness of this strategy has been demonstrated through the analysis of various published datasets, encompassing a range of fault scenarios, and its application in a real building equipped with fan coil units.

Furthermore, the impact of Principal Component Analysis (PCA) as a dimensionality reduction tool on the clustering algorithm's performance was investigated. It was observed that PCA generally had a detrimental effect on the clustering outcomes, particularly in scenarios involving multiple or subtle faults. This influence, however, may vary with datasets characterized by different levels of noise and variation.

One noted limitation of the proposed framework is the manual necessity in differentiating faulty and normal data post-clustering. Commonly, smaller clusters are assumed to represent faulty data, given the relative rarity of faults in datasets. However, this assumption is not universally applicable and fails to hold in all instances. Additionally, the current method is not optimized for detecting minor faults and lacks the capability for automated fault diagnosis. It also falls short in identifying simultaneous faults.

The process of threshold adjustment, crucial for cluster detection, requires careful consideration. Inconsistent data fluctuations can lead to false alarms by misidentifying significant clusters as faults. Future studies should focus on automating threshold adjustment and validating cluster extractions. By accumulating a limited dataset of known faults, this method could be extended to semi-supervised learning applications, enhancing fault detection within HVAC systems. Moreover, the availability of fault-free data could significantly contribute to new event detection strategies.

## 9 Acknowledgments

This research has been funded by the Natural Science and Engineering Research Council (NSERC) Alliance Grant (ALLRP- 544569–19) and FuseForward.

## 10 References

[1] S. Katipamula and M. R. Brambley, "Methods for fault detection, diagnostics, and prognostics for building systems—a review, part I," *Hvac&R Research,* vol. 11, no. 1, pp. 3-25, 2005.
[2] S. Katipamula and M. R. Brambley, "Methods for fault detection, diagnostics, and prognostics for building systems—A review, part II," *Hvac&R Research,* vol. 11, no. 2, pp. 169-187, 2005.
[3] M. S. Mirnaghi and F. Haghighat, "Fault detection and diagnosis of large-scale HVAC systems in buildings using data-driven methods: A comprehensive review," *Energy and Buildings,* vol. 229, p. 110492, 2020.
[4] G. Hu, T. Zhou, and Q. Liu, "Data-driven machine learning for fault detection and diagnosis in nuclear power plants: A review," *Frontiers in Energy Research,* vol. 9, p. 663296, 2021.
[5] H. Yang, T. Zhang, H. Li, D. Woradechjumroen, and X. Liu, "HVAC equipment, unitary: Fault detection and diagnosis," *Encyclopedia of Energy Engineering and Technology, 2nd ed.; CRC Press: Boca Raton, FL, USA,* pp. 854-864, 2014.
[6] J. Chen, L. Zhang, Y. Li, Y. Shi, X. Gao, and Y. Hu, "A review of computing-based automated fault detection and diagnosis of heating, ventilation and air conditioning systems," *Renewable and Sustainable Energy Reviews,* vol. 161, p. 112395, 2022.
[7] T. Maile, V. Bazjanac, and M. Fischer, "A method to compare simulated and measured data to assess building energy performance," *Building and Environment,* vol. 56, pp. 241-251, 2012.
[8] S. U. Lee, F. L. Painter, and D. E. Claridge, "Whole-Building Commercial HVAC System Simulation for Use in Energy Consumption Fault Detection," *Ashrae Transactions,* vol. 113, no. 2, 2007.
[9] A. Ranade, G. Provan, A. E.-D. Mady, and D. O'Sullivan, "A computationally efficient method for fault diagnosis of fan-coil unit terminals in building Heating Ventilation and Air Conditioning systems," *Journal of Building Engineering,* vol. 27, p. 100955, 2020.
[10] B. Dong, Z. O'Neill, and Z. Li, "A BIM-enabled information infrastructure for building energy Fault Detection and Diagnostics," *Automation in Construction,* vol. 44, pp. 197-211, 2014.
[11] Y. Zhao, J. Wen, F. Xiao, X. Yang, and S. Wang, "Diagnostic Bayesian networks for diagnosing air handling units faults–part I: Faults in dampers, fans, filters and sensors," *Applied Thermal Engineering,* vol. 111, pp. 1272-1286, 2017.
[12] K. Verbert, R. Babuška, and B. De Schutter, "Combining knowledge and historical data for system-level fault diagnosis of HVAC systems," *Engineering Applications of Artificial Intelligence,* vol. 59, pp. 260-273, 2017.
[13] J. Cui and S. Wang, "A model-based online fault detection and diagnosis strategy for centrifugal chiller systems," *International Journal of Thermal Sciences,* vol. 44, no. 10, pp. 986-999, 2005.

[14] H. Li and J. E. Braun, "A methodology for diagnosing multiple simultaneous faults in vapor-compression air conditioners," *HVAC&R Research,* vol. 13, no. 2, pp. 369-395, 2007.
[15] Y. Zhao, T. Li, X. Zhang, and C. Zhang, "Artificial intelligence-based fault detection and diagnosis methods for building energy systems: Advantages, challenges and the future," *Renewable and Sustainable Energy Reviews,* vol. 109, pp. 85-101, 2019.
[16] M. S. Piscitelli, D. M. Mazzarelli, and A. Capozzoli, "Enhancing operational performance of AHUs through an advanced fault detection and diagnosis process based on temporal association and decision rules," *Energy and Buildings,* vol. 226, p. 110369, 2020.
[17] M. Ester, H.-P. Kriegel, J. Sander, and X. Xu, "A density-based algorithm for discovering clusters in large spatial databases with noise," in *kdd*, 1996, vol. 96, no. 34, pp. 226-231.
[18] P.-N. Tan, M. Steinbach, and V. Kumar, *Introduction to data mining*. Pearson Education India, 2016.
[19] R. Yan, Z. Ma, Y. Zhao, and G. Kokogiannakis, "A decision tree based data-driven diagnostic strategy for air handling units," *Energy and Buildings,* vol. 133, pp. 37-45, 2016.
[20] D. Li, Y. Zhou, G. Hu, and C. J. Spanos, "Fault detection and diagnosis for building cooling system with a tree-structured learning method," *Energy and Buildings,* vol. 127, pp. 540-551, 2016.
[21] H. Han, X. Cui, Y. Fan, and H. Qing, "Least squares support vector machine (LS-SVM)-based chiller fault diagnosis using fault indicative features," *Applied Thermal Engineering,* vol. 154, pp. 540-547, 2019.
[22] K. Sun, G. Li, H. Chen, J. Liu, J. Li, and W. Hu, "A novel efficient SVM-based fault diagnosis method for multi-split air conditioning system's refrigerant charge fault amount," *Applied Thermal Engineering,* vol. 108, pp. 989-998, 2016.
[23] Y. Zhao, S. Wang, and F. Xiao, "A system-level incipient fault-detection method for HVAC systems," *HVAC&R Research,* vol. 19, no. 5, pp. 593-601, 2013.
[24] S. Miyata, J. Lim, Y. Akashi, Y. Kuwahara, and K. Tanaka, "Fault detection and diagnosis for heat source system using convolutional neural network with imaged faulty behavior data," *Science and Technology for the Built Environment,* vol. 26, no. 1, pp. 52-60, 2020.
[25] Y. Guo *et al.*, "Optimized neural network-based fault diagnosis strategy for VRF system in heating mode using data mining," *Applied Thermal Engineering,* vol. 125, pp. 1402-1413, 2017.
[26] Y. H. Eom, J. W. Yoo, S. B. Hong, and M. S. Kim, "Refrigerant charge fault detection method of air source heat pump system using convolutional neural network for energy saving," *Energy,* vol. 187, p. 115877, 2019.
[27] J. Liu, J. Liu, H. Chen, Y. Yuan, Z. Li, and R. Huang, "Energy diagnosis of variable refrigerant flow (VRF) systems: Data mining technique and statistical quality control approach," *Energy and Buildings,* vol. 175, pp. 148-162, 2018.
[28] Y. Zhao, S. Wang, and F. Xiao, "A statistical fault detection and diagnosis method for centrifugal chillers based on exponentially-weighted moving average control charts and support vector regression," *Applied Thermal Engineering,* vol. 51, no. 1-2, pp. 560-572, 2013.
[29] Z. Chen *et al.*, "A review of data-driven fault detection and diagnostics for building HVAC systems," *Applied Energy,* vol. 339, p. 121030, 2023.
[30] K. Yan, C. Zhong, Z. Ji, and J. Huang, "Semi-supervised learning for early detection and diagnosis of various air handling unit faults," *Energy and Buildings,* vol. 181, pp. 75-83, 2018.
[31] C. Fan, Y. Liu, X. Liu, Y. Sun, and J. Wang, "A study on semi-supervised learning in enhancing performance of AHU unseen fault detection with limited labeled data," *Sustainable Cities and Society,* vol. 70, p. 102874, 2021.
[32] Z. Du, B. Fan, X. Jin, and J. Chi, "Fault detection and diagnosis for buildings and HVAC systems using combined neural networks and subtractive clustering analysis," *Building and Environment,* vol. 73, pp. 1-11, 2014.

[33] B. Narayanaswamy, B. Balaji, R. Gupta, and Y. Agarwal, "Data driven investigation of faults in HVAC systems with model, cluster and compare (MCC)," in *Proceedings of the 1st ACM Conference on Embedded Systems for Energy-Efficient Buildings*, 2014, pp. 50-59.

[34] E. Novikova, M. Bestuzhev, and A. Shorov, "The visualization-driven approach to the analysis of the HVAC data," in *Intelligent Distributed Computing XIII*, 2020: Springer, pp. 547-552.

[35] R. Yan, Z. Ma, G. Kokogiannakis, and Y. Zhao, "A sensor fault detection strategy for air handling units using cluster analysis," *Automation in Construction,* vol. 70, pp. 77-88, 2016.

[36] J. Ploennigs, B. Chen, A. Schumann, and N. Brady, "Exploiting generalized additive models for diagnosing abnormal energy use in buildings," in *Proceedings of the 5th ACM Workshop on Embedded Systems For Energy-Efficient Buildings*, 2013, pp. 1-8.

[37] J. E. Seem, "Using intelligent data analysis to detect abnormal energy consumption in buildings," *Energy and buildings,* vol. 39, no. 1, pp. 52-58, 2007.

[38] A. Srivastav, A. Tewari, and B. Dong, "Baseline building energy modeling and localized uncertainty quantification using Gaussian mixture models," *Energy and Buildings,* vol. 65, pp. 438-447, 2013.

[39] Y. Yu, D. Woradechjumroen, and D. Yu, "A review of fault detection and diagnosis methodologies on air-handling units," *Energy and Buildings,* vol. 82, pp. 550-562, 2014.

[40] Y. Chen, G. Lin, Z. Chen, J. Wen, and J. Granderson, "A simulation-based evaluation of fan coil unit fault effects," *Energy and Buildings,* vol. 263, p. 112041, 2022.

[41] T. Caliński and J. Harabasz, "A dendrite method for cluster analysis," *Communications in Statistics-theory and Methods,* vol. 3, no. 1, pp. 1-27, 1974.

[42] M. Ankerst, M. M. Breunig, H.-P. Kriegel, and J. Sander, "OPTICS: Ordering points to identify the clustering structure," *ACM Sigmod record,* vol. 28, no. 2, pp. 49-60, 1999.

[43] H. P. Kriegel, P. Kröger, J. Sander, and A. Zimek, "Density-based clustering," *Wiley interdisciplinary reviews: data mining and knowledge discovery,* vol. 1, no. 3, pp. 231-240, 2011.

[44] S. Wang and F. Xiao, "AHU sensor fault diagnosis using principal component analysis method," *Energy and Buildings,* vol. 36, no. 2, pp. 147-160, 2004.

[45] J. Granderson *et al.*, "A labeled dataset for building HVAC systems operating in faulted and fault-free states," *Scientific Data,* vol. 10, no. 1, p. 342, 2023.

[46] J. Granderson *et al.*, "LBNL fault detection and diagnostics datasets," DOE Open Energy Data Initiative (OEDI); Lawrence Berkeley National Lab.(LBNL …, 2022.

[47] H. B. Gunay and Z. Shi, "Cluster analysis-based anomaly detection in building automation systems," *Energy and Buildings,* vol. 228, p. 110445, 2020.